\documentclass[sigconf]{acmart}

\usepackage{booktabs} 
\usepackage{xspace}
\usepackage{color}
\usepackage{url}
\usepackage[ruled]{algorithm2e} 
\usepackage{multirow}
\usepackage{tikz}
\usetikzlibrary{decorations.pathreplacing,calc}
\newcommand{\tikzmark}[1]{\tikz[overlay,remember picture] \node (#1) {};}

\newcommand*{\AddNote}[4]{%
    \begin{tikzpicture}[overlay, remember picture]
        \draw [decoration={brace,amplitude=0.5em},decorate,ultra thick,black]
            ($(#3)!(#1.north)!($(#3)-(0,1)$)$) --  
            ($(#3)!(#2.south)!($(#3)-(0,1)$)$)
                node [align=left, text width=4cm, pos=0.5, anchor=west] {{\bf #4}};
    \end{tikzpicture}
}%
\newcommand\tab[1][0.3cm]{\hspace*{#1}}
\makeatletter
\newcommand*{\bdiv}{%
  \nonscript\mskip-\medmuskip\mkern5mu%
  \mathbin{\operator@font div}\penalty900\mkern5mu%
  \nonscript\mskip-\medmuskip
}
\makeatother

\newcommand{\CPTWOK}{{\tt CP2K}\xspace}
\newcommand{\DBCSR}{{\tt DBCSR}\xspace}

\newcommand{\IE}{{i.\,e.}\xspace}

\DeclareMathOperator{\lcm}{lcm}

\begin{document}
\title[Increasing the Efficiency of Sparse Matrix-Matrix Multiplication]{Increasing the Efficiency of Sparse Matrix-Matrix Multiplication with a 2.5D Algorithm and One-Sided MPI}

\author{Alfio Lazzaro}
\orcid{ }
\affiliation{%
  \institution{University of Z\"urich, Department of Chemistry C}
  \streetaddress{Winterthurerstrasse 190}
  \city{Z\"urich} 
  \country{Switzerland} 
  \postcode{8057}
}
\email{alfio.lazzaro@chem.uzh.ch}

\author{Joost VandeVondele}
\affiliation{%
  \institution{CSCS}
  \streetaddress{Wolfgang-Pauli-Strasse 27}
  \city{Z\"urich} 
  \country{Switzerland} 
  \postcode{8093}
}
\email{joost.vandevondele@cscs.ch}

\author{J\"urg Hutter}
\affiliation{%
  \institution{University of Z\"urich, Department of Chemistry C}
  \streetaddress{Winterthurerstrasse 190}
  \city{Z\"urich} 
  \country{Switzerland} 
  \postcode{8057}
}
\email{hutter@chem.uzh.ch}

\author{Ole Sch\"utt}
\affiliation{%
  \institution{Empa}
  \streetaddress{\"Uberlandstrasse 129}
  \city{D\"ubendorf} 
  \country{Switzerland}
  \postcode{8600}
}
\email{ole.schuett@empa.ch}

\begin{abstract}
Matrix-matrix multiplication is a basic operation in linear algebra and an essential building block for a wide range of algorithms in various scientific fields. Theory and implementation for the dense, square matrix case are well-developed. If matrices are sparse, with application-specific sparsity patterns, the optimal implementation remains an open question. Here, we explore the performance of communication reducing 2.5D algorithms and one-sided MPI communication in the context of linear scaling electronic structure theory. In particular, we extend the \DBCSR sparse matrix library, which is the basic building block for linear scaling electronic structure theory and low scaling correlated methods in \CPTWOK.
The library is specifically designed to efficiently perform block-sparse matrix-matrix multiplication of matrices with a relatively large occupation. Here, we compare the performance of the original implementation based on Cannon's algorithm and MPI point-to-point communication, with an implementation based on MPI one-sided communications (RMA), in both a 2D and a 2.5D approach. The 2.5D approach trades memory and auxiliary operations for reduced communication, which can lead to a speedup if communication is dominant. The 2.5D algorithm is somewhat easier to implement with one-sided communications. A detailed description of the implementation is provided, also for non ideal processor topologies, since this is important for actual applications. Given the importance of the precise sparsity pattern, and even the actual matrix data, which decides the effective fill-in upon multiplication, the tests are performed within the \CPTWOK package with application benchmarks. Results show a substantial boost in performance for the RMA based 2.5D algorithm, up to 1.80x, which is observed to increase with the number of involved processes in the parallelization.
\end{abstract}

\copyrightyear{2017}
\acmYear{2017}
\setcopyright{acmlicensed}
\acmConference{PASC '17}{June 26-28, 2017}{Lugano, Switzerland}
\acmPrice{15.00}
\acmDOI{http://dx.doi.org/10.1145/3093172.3093228}
\acmISBN{978-1-4503-5062-4/17/06}


%
%



\maketitle

\section{Introduction}

Multiplication of two sparse matrices (SpGEMM) is a key operation in the the simulation of the electronic structure of systems containing thousands of atoms and electrons~\cite{joost1M}. Examples of such systems include electronic devices, complex interfaces, macromolecules or large disordered systems, with applications in the fields of renewable energy and electronics. The theory that enables such studies is linear scaling Density Functional Theory (DFT)~\cite{LS}. 
In the atomistic simulations package \CPTWOK~\cite{cp2k}, the linear scaling DFT implementation exploits the fact that operators in a localized atomic basis are sparse. The matrices have several thousands of non-zero elements per row and a priori unknown sparsity patterns.
This includes the Kohn-Sham matrix ($H$), the overlap matrix ($S$), and the density matrix ($P$). If $P$ can be computed from $H$ and $S$ without explicit reference to eigenvectors of $(H,S)$,
the traditional, cubically scaling approach of diagonalization can be avoided, and potentially replaced with a linear scaling method. 
We can obtain the density matrix from its functional definition 
\begin{equation}
P = \frac{1}{2} (I - {\rm sign}(S^{-1} H - \mu I))S^{-1},
\end{equation}
where $I$ is the identity matrix and $\mu$ is the chemical
potential. The matrix sign function is defined as
\begin{equation}
{\rm sign}(A) = A(A^2)^{-1/2}.
\end{equation}
Eigenvectors of $A$ are eigenvectors of ${\rm sign}(A)$,
with the eigenvalues of ${\rm sign}(A)$ being $-1$ or $+1$ for negative or positive eigenvalues of $A$, respectively. In its simplest form, we exploit the fact that the matrix sign function can be computed with a simple iterative scheme based only on matrix multiplications (two multiplications per iteration)
\begin{equation}
X_{n+1}=\frac{1}{2}X_n(3I-X^2_n),
\label{eq:iter}
\end{equation}
where $X_{n+1}$ converges to the sign of $X_0$~\cite{Higham:2008:FM}. 
A linear scaling cost results from the fact that all matrix operations are performed on sparse matrices, which have a number of non-zero entries that scale linearly with system size.
A particular characteristic of the involved matrices is that they are block-sparse, instead of element-wise sparse, where the dimensions of the blocks depend on the atomic kinds present in the studied systems. In order to retain sparsity during the iterations of the above
algorithm, a filtering multiplication is employed in two phases: on-the-fly during the product of two atomic blocks and after each multiplication to ignore small matrix elements~\cite{joost1M}.

In general, SpGEMM accounts for more than 80\% of the total runtime. The computational cost depends strongly on the evolution of the sparsity during the iterations,
which in turn depends on the chemical properties of the system studied, the precise algorithm employed, the system size, and the required accuracy~\cite{joost1M}.
The highly optimized sparse linear algebra library \DBCSR (Distributed Block Compressed Sparse Row) has been specifically designed to efficiently perform block-sparse matrix-matrix multiplications~\cite{dbcsr, ole}. It is parallelized using MPI and OpenMP, and can exploit GPU accelerators by means of CUDA and OpenCL. Prior to this work, Cannon's algorithm was used to parallelize the matrix-matrix multiplication~\cite{cannon}, using MPI point-to-point communications. 

In this paper we present a novel approach to MPI parallelization of the \DBCSR sparse matrix-matrix multiplication algorithm based on the communication-reducing 2.5D algorithm~\cite{25d}. This algorithm speeds up the execution by reducing the volume of transferred data with respect to the original \DBCSR algorithm. The implementation is based on MPI one-sided communications. 
For the next generation of computing systems, the number of nodes is expected to increase at an higher rate than the network performance.
Therefore, reducing time for communications by reducing the volume of exchanged data is crucial. The paper is organized as follows.
In the section~\ref{sec:DBCSR} we introduce \DBCSR, focusing on the MPI parallelization, followed by the description of the new communication-reducing implementation in section~\ref{sec:RMA}.
Finally, in section~\ref{sec:results} the performance  results  for  representative \CPTWOK  benchmarks are reported.
The source code of the implementation as well as the benchmarks are freely available to encourage reproducing these results.

\subsection{Related work}
The classical serial SpGEMM algorithm was first described by Gustavson~\cite{Gustavson:1978:TFA:355791.355796}.
The parallel implementation in a distributed memory system presents several challenges, such as load-balance and 
communication costs relative to arithmetic operations~\cite{10.1109/ICPP.2008.45}. To tackle these problems, in the past decade several methods have been presented. When there is a domain related knowledge about the input matrix sparsity structure, this information can be used to improve performance~\cite{ibm}. Similarly, by employing an initially~\cite{Ballard:2016:HPS:3012407.3015144} computed symbolic SpGEMM, a prior knowledge of a fixed pattern of the sparsity structure of the output matrix can be exploited. \DBCSR considers the general case
where a priori knowledge of the input and output matrix sparsity is not employed, and is aimed at delivering good performance if the matrix contains many non-zeros per row or is nearly dense. It uses a random permutation of the rows and columns of the matrix to achieve a good average load-balance (see section~\ref{sec:DBCSR}). Consequently, the data and the corresponding operations are statically distributed across processes in the same way as for dense matrices, and existing algorithms for dense matrix-matrix multiplications (e.g. ~\cite{Ballard:2013:COP:2486159.2486196}) can be adopted and refined for the sparse case. In particular, the development of so-called 3D algorithms for dense matrix multiplication allows to reduce the communication costs relative to arithmetic operations~\cite{3d_dense, Agarwal95athree-dimensional, IRONY20041017, 25d}.
Recently, Azad~{\it et al.} have reported the implementation of a multi-threaded 3D SpGEMM algorithm and presented performance results for applications in several fields~\cite{3d_algos}. However, this implementation does not consider block-sparse matrices and is not enabled for hybrid multi-cores CPU and GPU systems, both aspects are key and heavily optimized~\cite{ole} in \DBCSR. 
Rubensson and Rudberg reported a parallel implementation based on a different approach: instead of randomization, data locality is employed in their implementation for reducing communications, while the
mapping of data and work to physical resources is performed dynamically during the calculation~\cite{rubensson_rudberg}.
Like \DBCSR, this implementation is able to work effectively with block-sparse matrices and runs on hybrid multi-cores CPU and GPU systems. This implementation does not employ optimized libraries for the small block multiplications, nor on-the-fly filtering. 
To the best of our knowledge, the new \DBCSR implementation described in this paper is the only library for SpGEMM based on MPI one-sided communications that runs on hybrid multi-cores CPU and GPU systems.

\section{{\large \textbf{\DBCSR}} Library}
\label{sec:DBCSR}

\DBCSR is written in Fortran and is freely available under the GPL license from public repositories at {\tt sourceforge.net} and {\tt github.com} repositories.
It has a flexible and powerful API that can be explored online~\cite{cp2k_api}.
The \DBCSR library is designed to be highly efficient in the limit of high occupation ($>$10\%) of block sparse matrices, typically several thousands of non-zeros
per row, but no specific sparsity structure.
\DBCSR matrices are stored in the blocked compressed sparse row (CSR) format distributed over a two-dimensional grid of $P$ processes (each process holds a \emph{panel} of the matrix). Individual matrix elements are grouped into blocks by rows and columns. The blocked rows and columns form a grid of blocks. Randomized permutation of rows and columns is used to obtain a good average load-balance, and a static decomposition of the data across processes. Note that, sparsity patterns of the three matrices involved in multiplications are not identical. The result matrix has a non-fixed sparsity pattern, that is determined as the result of the computation, with blocks that are smaller than a given threshold removed after or skipped during the multiplication process.

In the original \DBCSR, inter-process communication is based on Cannon's algorithm~\cite{cannon}, the data of the matrix multiplication $C = C + A\cdot B$ is decomposed such that $C$ panels are always local, \IE each process computes a given $C$ panel, which is thus not communicated. The algorithm is generalized to an arbitrary two-dimensional process grid $P = P_R \cdot P_C$, 
with $P_R$ process rows and $P_C$ process columns. To achieve that, an additional virtual process dimension, $V = \lcm(P_R, P_C)$ is defined to match matrices in multiplication, \IE matrix $A$ is mapped to a $(P_R \cdot V)$ virtual process grid and matrix $B$ is mapped to a $(V \cdot P_C)$ process grid.
After a pre-shift of the data following Cannon's scheme on the virtual process grid (based on MPI point-to-point blocking calls), the algorithm performs $V$ steps for each multiplication (\emph{ticks}). 
The number of ticks is minimal when $P_R==P_C$, which implies $V=\mathcal{O}(\sqrt{P})$, therefore a square number of processes is optimal, or at least when $P_R$ and $P_C$ have most of their factors in common. 
A local multiplication and a data transfer for $A$ and $B$ panels between virtual neighboring processes are required, allowing a natural overlap between communications and computation. The volume of communicated data by each process scales as $\mathcal{O} (1/\sqrt{P})$. Transfer of all data between neighboring processes in the 2D grid (row-wise and column-wise shifts for $A$ and $B$ matrices, respectively) is performed using non-blocking MPI calls ({\tt mpi\_isend / mpi\_irecv}) for each tick. The multiplication starts as soon as all data have arrived at the destination process (by using a \linebreak {\tt mpi\_waitall} call). In total, the implementation requires 4 temporary buffers, meaning 2 buffers each (communication and computation) for matrices $A$ and $B$. A schematic description of the algorithm is presented in Algorithm~\ref{alg:cannon}.

\begin{algorithm}[t]
\SetAlgoNoLine
\KwIn{$A$, $B$ and $C$ matrices distributed over $(P_R \cdot P_C)$ process grid so that $P_{ij}$ process owns $A_{ij}$, $B_{ij}$ and $C_{ij}$ panels ($i=0,\dots,P_R-1$; $j=0,\dots,P_C-1$).}
\KwOut{$C= C+ A \cdot B$ matrix distributed over $(P_R \cdot P_C)$ process grid so that $P_{ij}$ process owns $C_{ij}$ panel.}
\tcc{do in parallel with all $P_{ij}$ process}
Allocate $A$ and $B$ buffers for communication ({\it comm}) and computation ({\it comp}); \\
Row-wise pre-shift by $i$ processes of the $A$ matrix, pointed by $A_{comp}$; \\
Column-wise pre-shift by $j$ processes of the $B$ matrix, pointed by $B_{comp}$; \\
$nticks = \lcm(P_R, P_C)$;\\
\For{$itick=1 \to nticks$}{
    \If{$itick>1$}{
        Ensures communication from previous iteration is complete, \IE new data has arrived in {\it comp} buffers and {\it comm} buffer data has been sent (call to {\tt mpi\_waitall()});
        }
    \If{$itick<nticks$}{
        Column and row shifts between neighbor processes, data is sent from the {\it comp} buffers and received into the {\it comm} buffers (call to {\tt mpi\_irecv()} and {\tt mpi\_isend()});
    }
    $C_{ij}= C_{ij}+A_{comp}\cdot B_{comp}$; \\
    Swap {\it comm} and {\it comp} buffer pointers;
}
\caption{Original \DBCSR matrix-matrix multiplication algorithm based on Cannon's algorithm and MPI point-to-point communications.}
\label{alg:cannon}
\end{algorithm}

The local multiplication consists of multiplications of matrix blocks.
They are organized in batches of block-wise small matrix-matrix multiplications that are processed by the CPU or alternatively by a GPU~\cite{ole}.
A filtering procedure is applied on the multiplication (on-the-fly filtering) of the blocks so that only blocks for which the product of their norms exceeds a given threshold will be actually multiplied. This filtering increases sparsity but also avoids performing calculations that fall below the filtering threshold, which results in a significant speed-up of the entire operation~\cite{joost1M}.
Multiple batches can be computed in parallel on the CPU by means of OpenMP threads. 
Processing these batches has to be highly efficient. For this reason specific libraries were developed, that outperform vendor basic linear algebra libraries (BLAS)\cite{dcse3, libxsmm, ole}. 

\section{Communication-reducing {\large \textbf{\DBCSR}} implementation}
\label{sec:RMA}

In the new implementation of the \DBCSR library, based on a 2.5D algorithm,  
panels for  matrices $A$ and $B$, which are distributed over the $(P_R \cdot P_C)$ process grid following the same scheme described in the section~\ref{sec:DBCSR},
are first copied in two buffers. These buffers are read-only within each multiplication, and reused between multiplications, by reallocating them only if the required size is larger than their actual size. 
The panels are used for creating MPI windows. To avoid the unnecessary blocking collective operation of creating and destroying the windows for each multiplication (\IE two collectives per $A$ and $B$), an {\tt mpi\_iallreduce} operation is executed beforehand to check if any of the memory pool in the windows requires a reallocation. This {\tt mpi\_iallreduce} operation overlaps with the initialization of the multiplication, so that it has a negligible impact in the overall execution time ($<\%1$). In this way we are able to limit the number of blocking collectives since the sparsity of the matrix will stabilize after a few initial iterations, and a maximum size of the buffers will be reached. Tests show that this optimization can give up to $5\%$ overall speedup, mainly due to reduced synchronization.

The $C$ matrix is distributed over the $(P_R \cdot P_C)$ process grid. 
Then the computation to obtain each resulting $C$ panel is split among $L$ processes, where $L$ represents the additional dimension when compared to the Cannon's algorithm. As $L$ is usually small, gradually transitioning from 2D to fully 3D, it is referred to as 2.5D~\cite{25d}. 
In turn, each process computes the partial multiplications for $L$ different $C$ panels. 
For $L>1$, at the end of a multiplication, these panels have to be communicated to the corresponding process in the 2D grid and accumulated to obtain the final resulting $C$ panel, \IE $L-1$ communications and accumulations, where we exclude the panel that already resides on the final process.

We distinguish two cases for the values of $L$, in addition to the trivial value $L=1$:
\begin{itemize}
    \item Non-square topology ($P_R \neq P_C$). Assuming $mn=\min(P_R,P_C)$ and $mx=\max(P_R,P_C)$, we require $mx$ to be an integer multiple of $mn$ and $mx\leq mn^2$. If so, we set $L=mx/mn$ and the corresponding 3D topology becomes 
    \begin{equation}
        mn \cdot \frac{mx}{L} \cdot L,
    \end{equation}
    \IE only the maximum dimension is scaled by $L$.
    \item Square topology ($P_R == P_C$). In this case $L$ can assume any square integer value such that $P_R$ has to be an integer multiple of $\sqrt{L}$. The 3D topology becomes
    \begin{equation}
        \frac{P_R}{\sqrt{L}} \cdot \frac{P_C}{\sqrt{L}} \cdot L,
    \end{equation}
    where both $P_R$ and $P_C$ are scaled by $\sqrt{L}$.
\end{itemize}
As a direct consequence of these definitions, the value of $L$ is such that $P/L$ is a square number. 

The communications of $A$ and $B$ panels are based on RMA passive target~\cite{UsingAdvancedMPI}, using MPI one-sided calls ({\tt mpi\_rget}), by always accessing the data in the initial positions in the 2D processes grid  during the entire multiplication operation. Which means, it does not require any pre-shift and subsequent communication between process neighbors. Furthermore, contrary to the algorithm described in Ref.~\cite{25d}, our algorithm does not employ any redistribution of the data in a 3D processes grid, but instead the 2D data partitioning is retained for performance, which is a consequence of the cost of such an operation in the presence of the sparsity, and one motivation for the use of RMA. MPI sub-communicators are used for the communications of the $A$ and $B$ panels between the corresponding MPI windows. The communications of the partial $C$ results employ MPI point-to-point communications ({\tt mpi\_isend / mpi\_irecv} calls), where MPI sub-communicators are specifically set for these communications.

Besides the two buffers used for the MPI windows, 
the number of temporary buffers for each process to store $A$, $B$ and $C$ panels are:
\begin{itemize}
    \item $L-1$ computation buffers for the partial result of the $C$ panels and a communication buffer used in their final accumulation (when $L>1$);
    \item in the case of square topology there are $\max(2,\sqrt{L})+2$ buffers for the $A$ and $B$ panels, while only $4$ buffers are needed for the non-square case.
\end{itemize}
In total, the implementation requires $6$ temporary buffers when $L==1$ (two more buffers than the \DBCSR implementation based on MPI point-to-point communications), $L+6$ buffers for the non-square topology, and
$L+\sqrt{L}+4$ for the square topology. Note that, as a result of multiplication of sparse matrices, the size of the $C$ panels ($S_C$) is in general larger than $A$ and $B$ panel sizes ($S_{A}$, $S_{B}$). Therefore the increases in memory footprint for the temporary buffers with respect to $L==1$ case become
\begin{equation}
\begin{aligned}
\textrm{non-square topology} & \rightarrow \frac{S_C}{3(S_A+S_B)}L+1, \\ 
\textrm{square topology} & \rightarrow \frac{S_C}{3(S_A+S_B)}L+\frac{\sqrt{L}+4}{6}.
\end{aligned}    
\label{eq:mem_increase}
\end{equation}
On the other hand, the amount of data communicated by each process for $A$ and $B$ panels reduces by a factor $\sqrt{L}$ as a consequence of to ability to reuse $A$ and $B$ panels in the evaluation of the $L$ panels of $C$. In this respect the algorithm trades memory for communications~\cite{McColl1999, Schatz2012ParallelMM}. In summary, the total amount of requested data by each process scales as
\begin{equation}
    \underbrace{\frac{V}{\sqrt{L}}(S_A+S_B)}_{A,\, B \textrm{ panels}} + 
    \underbrace{(L - 1)S_C}_{C \textrm{ panels}},
    \label{eq:comm25d}
\end{equation}
which leads to $\mathcal{O} (1/\sqrt{PL})$ scaling for the amount of communicated data, 
whereas the memory footprint and overhead increase by $\mathcal{O}(L)$. 
Thus the value for $L$ has to be tuned such that the contribution to the communications of the second term in Equation~\eqref{eq:comm25d} remains small and the memory increase (Equation~\eqref{eq:mem_increase}) is reasonable. Finally, the benefit from the 2.5D implementation becomes larger with higher number of processes.

Compared to the Cannon's algorithm, the number of ticks for the new implementation becomes $V/L$. 
As previously mentioned in section~\ref{sec:DBCSR}, also in this case a square topology is preferable since it leads to the minimum number of ticks.
Then, for each tick, each process performs the operations for the $L$ local $C$ panels, by considering the data panels in the 2D data layout.
The local multiplication for each panel will start as soon as all the data has arrived at the destination process (as a result of a {\tt mpi\_waitall} call). Furthermore, the communication of the $C$ panels starts during the last tick execution, allowing also overlap for this operation. Then, the accumulation to the local $C$ panels is organized such that the panel belonging to the local process, which does not need to be communicated, is used for the accumulation of the incoming partial results from the other $L-1$ processes. Note that the accumulation operations are entirely executed by the CPU, while the local $A\cdot B$ operations can be also executed by the GPU. 

A schematic description of the algorithm is presented in Algorithm~\ref{alg:25d}.

\begin{algorithm*}[p]
\SetAlgoNoLine
\KwIn{$A$, $B$ and $C$ matrices distributed over $(P_R \cdot P_C)$ process grid so that $P_{ij}$ process owns $A_{ij}$, $B_{ij}$ and $C_{ij}$ panels ($i=0,\dots,P_R-1$; $j=0,\dots,P_C-1$).}
\KwIn{$L$ value.}
\KwOut{$C= C+ A \cdot B$ matrix distributed over $(P_R \cdot P_C)$ process grid so that $P_{ij}$ process owns $C_{ij}$ panel.}
\tcc{do in parallel with all $P_{ij}$ process}
Initialization of read-only buffers and MPI windows for $A$ and $B$ matrices; \\ 
Check validity of $L$ for the $(P_R \cdot P_C)$ process grid, set $L=1$ if not valid; \\
$L_R = 1$; \tab $L_C = 1$; \tab $\textit{nbuffers}_A = 2$; \tikzmark{right_topo} \\
\tikzmark{top_nosquare}
\uIf{$P_R>P_C$} { 
    $L_R=L$; \\
}
\uElseIf{$P_R<P_C$} {
    $L_C=L$; \tikzmark{bottom_nosquare} \\ 
}
\Else {
    $L_R=\sqrt{L}$; \tab $L_C= L_R$; \tikzmark{top_square} \\
    $\textit{nbuffers}_A = \max(2, L_R)$; \tikzmark{bottom_square} \\
}
$side3D = \left(\max(P_R, P_C)\right) \bdiv \left(\max(L_R, L_C)\right)$; \tab $i3D = i \bdiv side3D$; \tab $j3D = j \bdiv side3D$; \tab $l = j3D\cdot L_R + i3D$; \tab $V = \lcm(P_R, P_C)$; \\ 
Allocate $\textit{nbuffers}_A$ buffers for $A$ ($A[0,\dots,\textit{nbuffers}_A-1]$) 
and 2 buffers for $B$ ($B[0,\dots,1]$); \\
Initialization of $L-1$ empty $C$ panels, pointed by $C[0,\dots,L_R-1][0,\dots,L_C-1]$ pointers, where $C[i3D][j3D]$ points to $C_{ij}$; \\
$lcomm_A = \textit{nbuffers}_A-1$; \tab $lcomm_B=1$; \\
\For{$t=0 \to V$}{
    \If{$t>0$}{
        Ensures communication (if any) from previous iteration is complete (call to {\tt mpi\_waitall()}); \tikzmark{right_loop} \\
    }
    \If{$t<V$}{
        \tikzmark{top_zero}
        \If{$(t \bmod L) == 0$}{
            $comm_A[0,\dots,L_R-1]=\mathbf{true}$; \tab $comm_B[0,\dots,L_C-1]=\mathbf{true}$; \tikzmark{bottom_zero}\\
        }
        $icomm3D = t\bmod L_R$; \tab $jcomm3D = \left(t\bdiv L_R\right) \bmod L_C$; \\
        $m = icomm3D \cdot side3D + i \bmod side3D$; \tab $n = jcomm3D \cdot side3D + j \bmod side3D$; \\
        \tikzmark{top_comm}
        \If{$comm_A[icomm3D]$}{
            $comm_A[icomm3D] = \mathbf{false}$; \tab $lcomm_A = (lcomm_A+1) \bmod \textit{nbuffers}_A$; \\
            $k= \left(j + ((i \cdot (V \bdiv P_R) + l + t) \cdot P_C) \bdiv V \right) \bmod P_C$; \\
            Request $A$ data from process $P_{mk}$ and put in buffer $A[lcomm_A]$ (call to {\tt mpi\_rget()}); \\
        }
        \If{$comm_B[jcomm3D]$}{
            $comm_B[jcomm3D] = \mathbf{false}$; \tab $lcomm_B = (lcomm_B+1) \bmod 2$; \\
            $k= \left(i + ((j \cdot (V \bdiv P_C) + l + t) \cdot P_R) \bdiv V \right) \bmod P_R$; \\
            Request $B$ data from process $P_{kn}$ and put in buffer $B[lcomm_B]$ (call to {\tt mpi\_rget()}); \tikzmark{bottom_comm}
        }
    }
    \If{$t>0$}{
        $C[icomp3D][jcomp3D] = C[icomp3D][jcomp3D] + A[lcomp_A] \cdot B[lcomp_B]$; \\
        \If{$((t>V-L) \; \mathbf{and} \; (L>1))$}{
            \tikzmark{top_red}
            $m = icomp3D \cdot side3D + i \bmod side3D$; \tab $n = jcomp3D \cdot side3D + j \bmod side3D$; \\
            Transfer $C[icomp3D][jcomp3D]$ partial result to process $P_{mn}$ for final reduction; \tikzmark{bottom_red} \\
        }
    }
    \tikzmark{top_swap}
    $icomp3D = icomm3D$; \tab $jcomp3D = jcomm3D$; \\
    \uIf{$((P_R==P_C) \; \mathbf{and} \; (L>1))$}{
        $lcomp_A = icomp3D$; \\
    }
    \Else {
        $lcomp_A = lcomm_A$; \\
    }
    $lcomp_B = lcomm_B$; \tikzmark{bottom_swap} \\
}
Finalize reduction of the partial results in $C_{ij}$;
\AddNote{top_nosquare}{bottom_nosquare}{right_topo}{\tab Non-square topology}
\AddNote{top_square}{bottom_square}{right_topo}{\tab Square topology}
\AddNote{top_zero}{bottom_zero}{right_loop}{\tab Start of a tick}
\AddNote{top_comm}{bottom_comm}{right_loop}{\tab Data reused within a tick}
\AddNote{top_red}{bottom_red}{right_loop}{\tab Last tick reduction}
\AddNote{top_swap}{bottom_swap}{right_loop}{\tab Swap $comm$ and $comp$\\ \tab buffers indices}
\caption{\DBCSR matrix-matrix multiplication algorithm based on 2.5D algorithm and MPI one-sided communications.}
\label{alg:25d}
\end{algorithm*}

\section{Performance results}
\label{sec:results}

We present the results of running \DBCSR with \CPTWOK benchmark applications, resulting in
matrices with different block sizes and occupation. This is important, as performance and scalability depend on these parameters. We compare the performance of the 2.5D implementation to the Cannon's algorithm implementation as available in the \DBCSR library, considering only the execution time of the \DBCSR multiplication part, and not any other application specific parts. Timings are taken as the average from 4 independent application runs, each consisting of tens of multiplications. Fluctuation are found to be less than 1\%.
Measured performance numbers are representatives of large-scale and long-running science runs of \CPTWOK for linear scaling calculations. Tests cover both strong and weak scaling.
Elements of the generated matrices are double precision floating point numbers.

All performance tests were carried out on Piz Daint, hosted at the Swiss National Supercomputing Centre (CSCS). At the time of the benchmark, Piz Daint was a CRAY XC30 machine with 5,272 compute nodes. Each node comprises a single socket 8-core Intel\textsuperscript{\textregistered} Xeon\textsuperscript{\textregistered}
E5-2670 (code-named Sandy Bridge) CPU at 2.6~GHz TDP frequency, one NIVIDA Tesla K20X, and 32~GB of RAM.
All CPU cores have Intel Turbo and Intel Hyper Threading Technology enabled. The
latter is not used in our benchmark runs, i. e. threads are pinned to cores individually.
 From an interconnect point of view, Piz Daint features CRAY's Aries network including
dedicated communication ASICs (shared by a group of four nodes) which are connected by a
Dragonfly high-radix network with a bisection bandwidth of 33 TB/s.
The code is compiled with GCC 4.9.2 against CRAY-MPICH 7.2.5 and CUDA 6.5.14. For the RMA runs 
DMAPP 7.0.1 is linked in, and the CRAY environment variable {\tt MPICH\_RMA\_OVER\_DMAPP=1} is set. Tests without linking DMAPP show an increase in execution time by a factor 2.4x on average.

A single MPI rank with 8 OpenMP threads is used on each node, which by itself reduces the amount of MPI communication. We found that this configuration gives the best performance with respect to a corresponding configuration with multiple ranks in a node (up to $10\%$ speedup) at the same total number of nodes.

\subsection{Strong scaling results}

We present the results of three benchmarks performed with \CPTWOK, representative of
matrices with a broad range of sparsity values:
\begin{itemize}
\item \verb+H2O-DFT-LS+: single-point 
energy calculation with linear  scaling DFT, consisting of 20,736 atoms in a 39 \AA\textsuperscript{3} box (6,912 water molecules in total). An LDA functional is used with a DZVP-MOLOPT basis set and a 300 Ry cutoff. This system implies matrices with medium sparsity (average occupancy 10\%).
\item \verb+S-E+: semi-empirical setup benchmark with 186,624 water molecules. This system implies matrices with large sparsity (average occupancy 0.05\%).
\item \verb+Dense+: fully occupied matrices, synthetic benchmark.
\end{itemize}

The block sizes, total number of rows/columns (all matrices are square), typical occupancy during the simulations, number of multiplications, and FLOPs executed by \DBCSR part only are reported in Table~\ref{table:DBCSR}.

\begin{table}
\centering
\caption{Block sizes, dimension of matrices (rows and columns), typical occupancy of the matrices, number of multiplications performed, and \DBCSR FLOPs for the three benchmarks. }
\label{table:DBCSR}
\centering
\begin{tabular}{lccc}
\hline
 & \texttt{H2O-DFT-LS} & \texttt{S-E} & \texttt{Dense} \\
\hline
Block sizes $(n \times n)$ & $23$ & $6$ & $32$ \\
\# Rows/columns &  $158,976$ & $1,119,744$ & $60,000$ \\
Occupancy range (\%) & $7 - 15$ &
                       $(4 - 6) \times 10^{-2}$ & 
                       $100$ \\
\# Multiplications & $193$ & $1198$ & $10$\\
\DBCSR FLOPs ($\times 10^{15}$)   & $4.038$ & $0.146$ & $4.320$ \\
\hline
\end{tabular}
\end{table}

\begin{figure}
\centering
\includegraphics[scale=0.46]{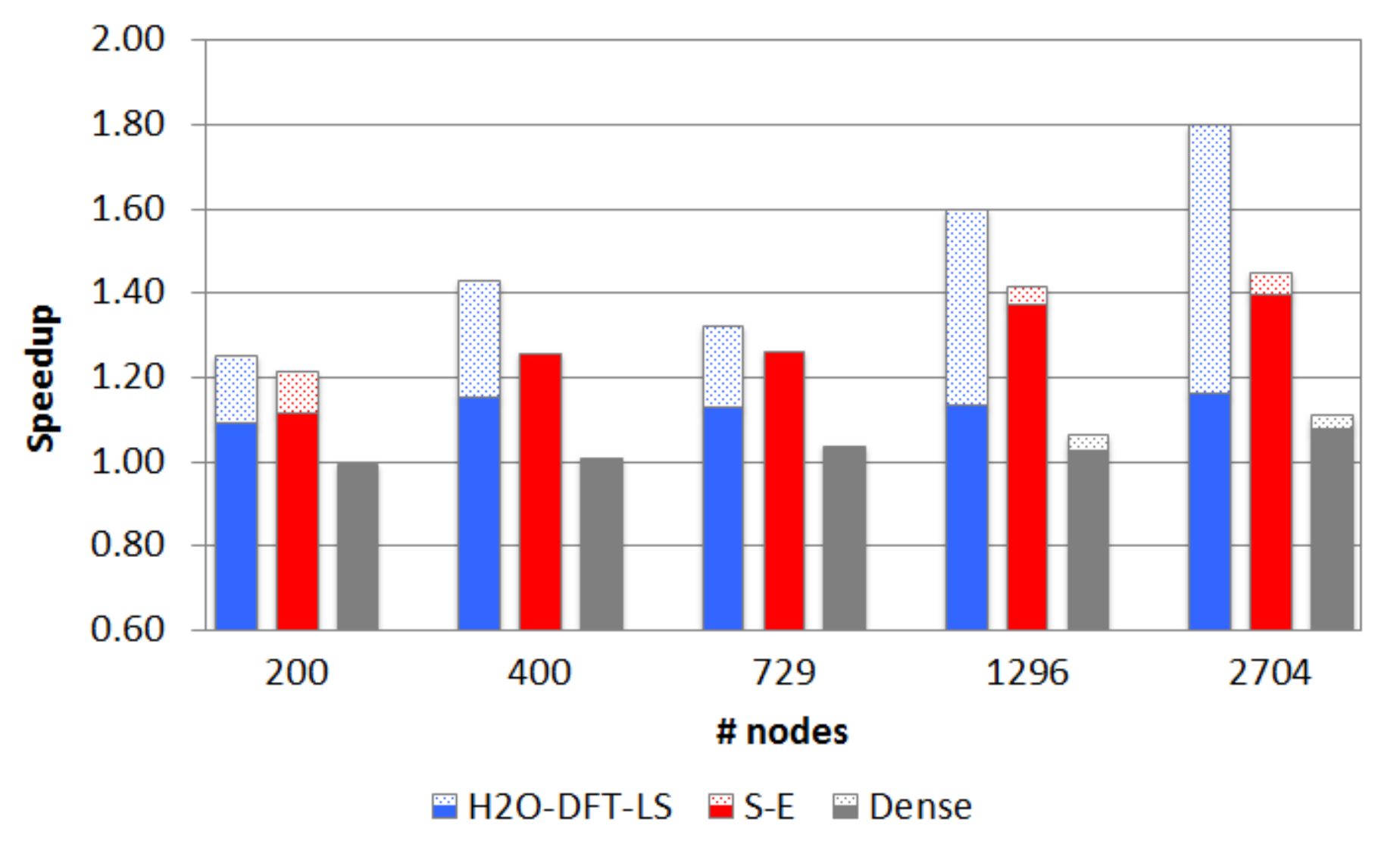}
\caption{Speedup values (higher is better) for the \DBCSR execution obtained from the ratios of the execution time of the point-to-point implementation with respect to the one-sided implementation with $L=1$ (solid filled bars on the front) and the fastest execution for any $L$ value (dot filled bars), varying the number of nodes in the strong scaling tests. The dot filled part gives the speedup when enabling the 2.5D algorithm with a value $L>1$ (where there is not such part visible, $L=1$ gives the fastest execution). 
For each number of nodes, the bars refer to (from left to right): \texttt{H2O-DFT-LS}, \texttt{S-E}, \texttt{Dense}.
}
\label{fig:strong_speedup}
\end{figure}

The \DBCSR multiplication execution time, the \DBCSR total amount of communicated data per process (average between all processes of the total exchanged data for $A$, $B$, and $C$ panels, see Equation~\eqref{eq:comm25d}), and the \CPTWOK peak memory footprint (maximum over all processes) for executions involving  different numbers of nodes for the two implementations with MPI point-to-point (PTP) and one-sided communications with different $L$ values (OS$L$) are reported in Table~\ref{table:strong_scaling}.
The reported peak memory footprint refers to the entire \CPTWOK application, \IE not just the \DBCSR part, therefore it shows the impact of changing $L$ in \DBCSR with respect to the entire \CPTWOK memory footprint (see Equation~\eqref{eq:mem_increase}).
The speedups of OS$L$ relative to PTP are shown in Figure~\ref{fig:strong_speedup}. 
Already the OS1 implementation gives faster executions than the original \DBCSR PTP implementation, with a speedup that increases with the number of nodes ranging from: 1.09x--1.16x for \texttt{H2O-DFT-LS}, 1.12x--1.40x for \texttt{S-E}, and 1.00x--1.08x for \texttt{Dense}. This speedup directly results from a reduction in the time spent in the {\tt mpi\_waitall} call that waits for the $A$ and $B$ panel communications to complete. This part of the algorithm is the limiting factor for the scalability of the \DBCSR multiplication execution. For instance, at 2704 nodes, which is the most prominent example for this effect, the fractions of this time with respect to the corresponding \DBCSR execution time for PTP and OS1 are, respectively: 57\% and 50\% (\texttt{H2O-DFT-LS}), 32\% and 5\% (\texttt{S-E}), 41\% and 37\% (\texttt{Dense}). 

\begin{figure}
\centering
\includegraphics[scale=0.46]{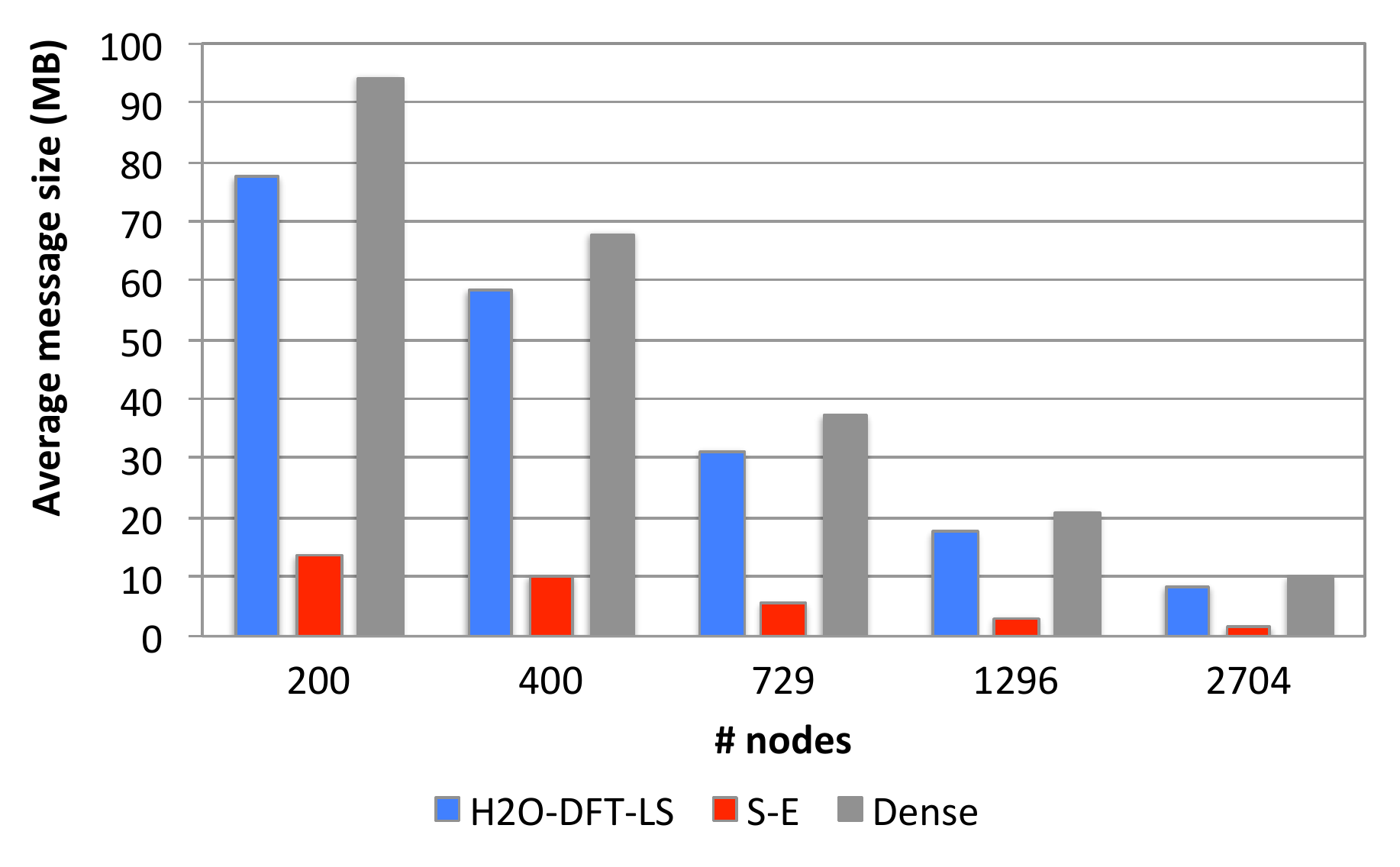}
\caption{Average message sizes (in MB) for the $A$ ($S_A$) and $B$ ($S_B$) panels exchange for the MPI point-to-point implementation or MPI one-sided implementation with $L=1$, varying the number of nodes in the strong scaling tests. For each benchmark, the values scale with the number of nodes in agreement with the expectations, 
assuming $S_A==2S_B$ for the virtual topology with 200 nodes and $S_A==S_B$ for the other cases with square topology.
For each number of nodes, the bars refer to (from left to right): \texttt{H2O-DFT-LS}, \texttt{S-E}, \texttt{Dense}.
}
\label{fig:size_strong}
\end{figure}

At this point, we remark that the timings are obtained from a \CPTWOK internal timing framework, annotating carefully the most important functions. Nevertheless, interpretation of the data is difficult, as the algorithm is largely asynchronous, both with computation on the GPU and with communication across the network. For example, the time spent in the {\tt mpi\_waitall} call is not the full communication time, but only the part that did not overlap with the other operations. In the future, tools that can analyze and visualize this at the scale of the experiments (several hundreds of nodes) will be useful. Nevertheless, we try to explain these timings with the following observations:
\begin{enumerate}
\item The one-sided algorithm implements a new communication scheme, which does not require the pre-shift of the data following Cannon's scheme.
\item The {\tt mpi\_waitall} completion in the point-to-point implementation requires synchronization on the sender and receiver processes, while the one-sided implementation has only synchronization on the receiver process. 
\item The seemingly large improvement in the \texttt{S-E} benchmark could be related to the average sizes of the exchanged messages in Figure~\ref{fig:size_strong}. The message sizes for the \texttt{S-E} benchmark are in average between 5.7x and 6.7x smaller than the other two benchmarks sizes. It is possible that the one-sided implementation performs especially better than point-to-point for smaller message sizes.
\end{enumerate}

\begin{table*}
\centering
\caption{\DBCSR multiplication execution time, \DBCSR total amount of communicated data per process, and \CPTWOK peak memory footprint for the strong scaling tests of the two implementations based on MPI point-to-point (PTP) and 2.5D one-sided with various $L$ values (OS$L$). \DBCSR total amount of communicated data per process is the average over all processes of the communicated data for $A$, $B$, and $C$ panels (see Equation~\eqref{eq:comm25d}) for all \DBCSR\ multiplications, where the values for PTP and OS1 scale in agreement with the expectations. 
The reported peak memory footprint refers to the entire \CPTWOK application, \IE not just the \DBCSR part, and is obtained from the maximum of the peak memory footprint over all processes.
}
\label{table:strong_scaling}
\centering
\begin{tabular}{cc|ccccc|ccccc|ccccc}
\cline{2-17} 
& \multirow{2}{*}{\# nodes}  & 
            \multicolumn{5}{c|}{\texttt{H2O-DFT-LS}} & 
            \multicolumn{5}{c|}{\texttt{S-E}} & 
            \multicolumn{5}{c}{\texttt{Dense}} \\
\cline{3-17} 
 &  & PTP & OS1 &  OS2 & OS4 &  OS9 & PTP &  OS1 & OS2 &  OS4 & OS9 &
 PTP & OS1 &  OS2 & OS4 &  OS9 \\
\hline
\multirow{5}{2cm}{\centering \DBCSR execution time (seconds)} 
& 200 & 325 & 298 & 260 & -- & -- & 558 & 500 & 459 &	-- & -- & 42.8 & 43.0 & 43.9 &	-- & -- \\
& 400 & 212	& 184 & -- & 148 & -- & 390 & 310 & -- & 310 &	-- & 22.1 & 21.9 & -- &	23.6 &	-- \\
& 729 & 155 & 137 & --	& -- & 117	& 310 & 246	& --	& --	& 314  & 13.3 & 13.3 & -- & -- & 15.5 \\ 
& 1296 & 136 & 120 & -- &	85 & 92 & 282 & 205 & -- & 199 & 254 & 11.2 & 10.9 & -- & 10.5 & 11.6 \\
& 2704 & 99 & 85 & -- &	55 & -- &	249 & 178 & -- & 172 & -- & 10.8 & 10.0 & -- & 9.7 & -- \\
\hline
\multirow{5}{2cm}{\centering \DBCSR total communicated data per process (GB)} 
& 200 & 640 & 640 & 491 & -- & -- & 856 & 856 & 630 &	-- & -- & 51 & 51 & 38 &	-- & -- \\
& 400 & 318	& 318 & -- & 228 & -- & 445 & 445 & -- & 286 &	-- & 26 & 26 & -- &	15 &	-- \\
& 729 & 236 & 236 & --	& -- & 145	& 329 & 329	& --	& --	& 200  & 20 & 20 & -- & -- & 10 \\ 
& 1296 & 177 &	177 & -- &	108 & 96 & 247 & 247 & -- & 140 & 125 & 15 & 15 & -- & 8 & 6 \\
& 2704 & 122 & 122 & -- &	70 & -- &	171 & 171 & -- & 93 & -- & 10 & 10 & -- & 5 & -- \\
\hline
\multirow{5}{2cm}{\centering \CPTWOK peak memory footprint (GB)} 
& 200 & 5.16 & 5.36 & 7.40 & -- & -- & 3.99 & 4.28 & 4.93 &	-- & -- & 2.54 & 2.68 & 3.41 &	-- & -- \\
& 400 & 3.20 & 3.26 & -- & 6.11 & -- & 3.00 & 3.10 & -- & 4.41 &	-- & 1.47 & 1.61 & -- &	2.52 &	-- \\
& 729 & 2.07 & 2.11 & --	& -- & 7.60	& 2.45 & 2.74	& --	& --	& 4.67  & 0.95 & 1.01 & -- & -- & 2.18 \\ 
& 1296 & 1.69 &	1.72 & -- &	3.33 & 6.11 & 2.19 & 2.34 & -- & 2.78 & 3.84 & 0.85 & 0.88 & -- & 1.19 & 1.75 \\
& 2704 & 0.95 & 0.98 & -- &	1.77 & -- &	2.10 & 2.17 & -- & 2.40 & -- & 0.57 & 0.58 & -- & 0.78 & -- \\
\hline
\end{tabular}
\end{table*}

\begin{figure}
\centering
\includegraphics[scale=0.46]{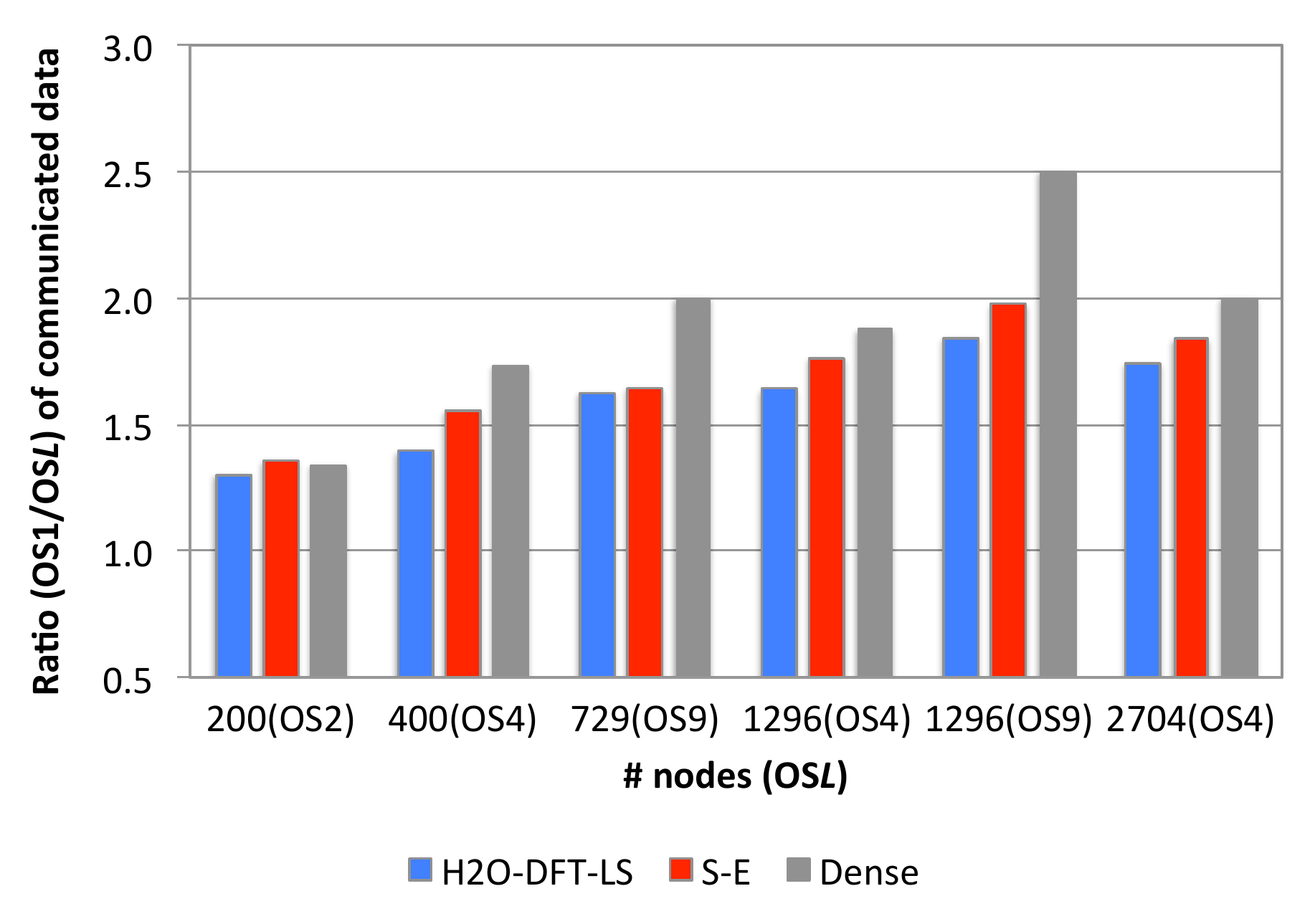}
\caption{Ratios of the \DBCSR total amount of communicated data per process between $L=1$ (OS1) and $L>1$ (OS$L$), varying the number of nodes and $L$ value in the strong scaling tests. For each number of nodes, the bars refer to (from left to right): \texttt{H2O-DFT-LS}, \texttt{S-E}, \texttt{Dense}.
}
\label{fig:size_layers}
\end{figure}

The OS$L$ implementation with $L>1$ allows to further improve the performance, especially for the communication-dominated \texttt{H2O-DFT-LS} benchmark (total speedup with respect to the PTP implementation at 2704 nodes is 1.80x). As expected, the boost increases when more nodes are involved.
The effect of introducing $L>1$ is the further reduction of the {\tt mpi\_waitall} timing for the $A$ and $B$ panels communications. However, this is offset by some overhead for handling partial $C$ panels and their communications and accumulations, executed by the CPU only (see Equation~\eqref{eq:comm25d}). The \texttt{Dense} benchmark, which has a large number of blocks to handle, is particularly affected by these factors: at 2704 nodes, the {\tt mpi\_waitall} timing for the $A$ and $B$ panels communications goes from 37\% for $L=1$ to 4\% for $L=4$ of the corresponding total \DBCSR execution time, but the overall speedup with respect to the PTP execution time is just 1.11x. For the \texttt{S-E} benchmark, the {\tt mpi\_waitall} timing for the $A$ and $B$ panels communications is already small with $L=1$ (5\%), therefore the effect of using $L>1$ is limited. We can analyze the effect of $L>1$ on the \DBCSR total amount of communicated data per process for the $A$, $B$, and $C$ panels (see values in Table~\ref{table:strong_scaling}). The ratios of these values between $L=1$ and $L>1$ are shown in Figure~\ref{fig:size_layers}. They are in agreement with the expectations obtained with the Equation~\eqref{eq:comm25d}, where the average ratios of sizes $S_C/S_{A,B}$ are: 2.7 for \texttt{H2O-DFT-LS}, 2.1 for \texttt{S-E}, and 1.0 for \texttt{Dense}. Although the volume of communicated data of $A$ and $B$ panels alone goes as $1/\sqrt{L}$, the reduction in overall volume is less as the contribution of $S_C$ becomes more dominant as $L$ increases. On the other hand, the overhead increases linearly with $L$, so that large values of $L$ are expected to pay off only at larger number of processes. 

Finally, we analyze the memory consumption. The \CPTWOK peak memory footprint values are reported in Table~\ref{table:strong_scaling}. A direct comparison with the ideal values as obtained from Equation~\eqref{eq:mem_increase} is only partially satisfactory. The measured peak memory usage is influenced by various other factors such as details on how the operating system allocates memory, or internal buffers of MPI and the rest of \CPTWOK that might vary depending on the run type. Nevertheless, we can understand the trends as the results vary between the benchmarks. The OS1 implementation requires on average 5\% more memory than the PTP implementation per process at the same number of nodes. Then, the memory footprint increases with $L>1$, also depending on $S_{A,B}$ and $S_C$ buffer sizes.
In particular, the \texttt{H2O-DFT-LS} benchmark shows the largest increment (average values with respect to OS1: 1.38x for OS2, 1.87x for OS4,
and 3.58x for OS9), which is directly related to the largest buffers sizes (average maximum values at 2704 nodes: $S_{A,B}=16$~MB, $S_C=45$~MB). The other two benchmarks present smaller increments (1.07x--2.07x) since their $S_{A,B}$ and $S_C$ are smaller (average maximum values at 2704 nodes: $S_{A,B}=3$~MB, $S_C=6$~MB for  \texttt{S-E}, $S_{A,B}=S_C=10$~MB for \texttt{Dense}).

To conclude this section, the 2.5D implementation based on MPI one-sided communications gives good performance in terms of execution time and memory consumption for reasonably small $L$ values.

\subsection{Weak scaling results}

\begin{figure*}
\centering
\includegraphics[scale=0.8]{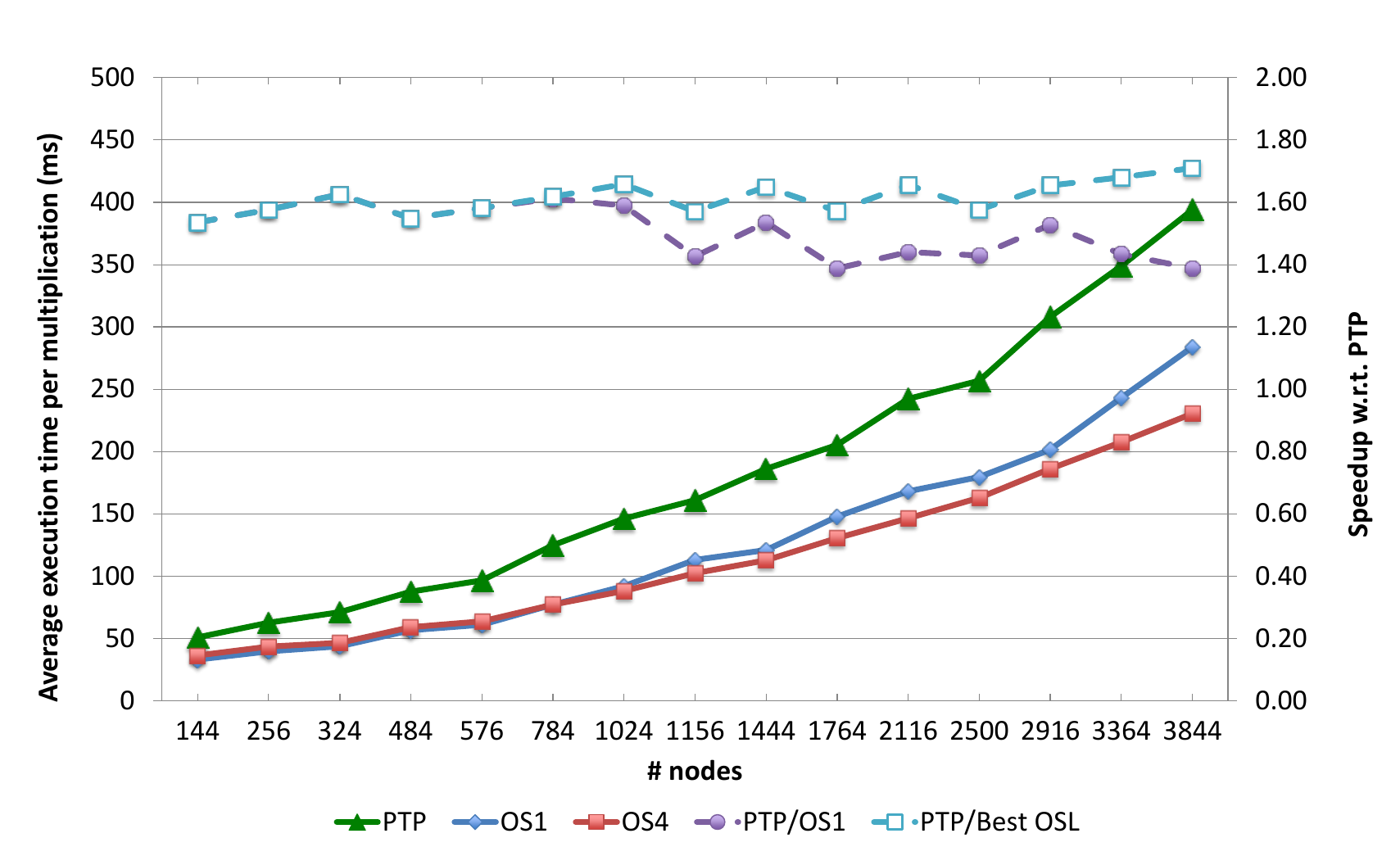}
\caption{Average execution time (in ms, left y-axis) of the \DBCSR multiplication part to perform a single multiplication for the MPI point-to-point implementation (PTP, solid line with triangular markers), MPI one-sided implementation with $L=1$ (OS1, solid line with rhomboid markers) and $L=4$ (OS4, solid line with square markers), varying the number of nodes in the weak scaling test with 617 multiplications. The ratios (right y-axis) of the PTP and OS1 (PTP/OS1, dashed line with circle markers) and PTP and the fastest OS implementation (PTP/Best OS$L$, dashed line with empty square markers) values are also reported. 
}
\label{fig:weak_scaling}
\end{figure*}

In this section, the performance of the weak scaling test is analyzed. The \texttt{S-E} benchmark is employed, fixing the number of water molecules to 76 per process, which leads to a constant amount of FLOPs and data per process, but growing communication and overhead costs. The sparsity of the matrices decreases linearly with the number of processes, from an average of 1.1\% on 144 nodes to 0.04\% on 3844 nodes. Based on the conclusion of the strong scaling benchmarks, we only considered square number of processes and $L=4$. The comparison of the execution time of the \DBCSR multiplication part for the PTP, OS1 and OS4 runs is shown in Figure~\ref{fig:weak_scaling}. Also in this case, OS1 outperforms PTP (the speedup is smaller with increasing number of nodes, because of the constant message sizes), and OS4 becomes beneficial for a large enough number of processes, reaching 1.7x with 3844 nodes.

\section{Conclusions}

The new \DBCSR implementation based on a 2.5D algorithm and MPI one-sided communications allows for reaching the same or better performance than the previous implementation based on Cannon's algorithm and MPI point-to-point communications. In the best case, a 1.80x speedup has been observed in our tests. The new communication scheme avoids the pre-shift of data that is required in Cannon's scheme and effectively trades memory for network bandwidth.

\section*{Acknowledgments}
This work was supported by grants from the Swiss National Supercomputing Centre (CSCS) under projects CH5 and D50 and received funding from the Swiss University Conference through the Platform for Advanced Scientific Computing (PASC). JV acknowledges financial support by the European Union FP7 in the form of an ERC Starting Grant under contract No. 277910.

\bibliographystyle{ACM-Reference-Format}
\bibliography{bibio} 


\begin{thebibliography}{00}


\ifx \showCODEN    \undefined \def \showCODEN     #1{\unskip}     \fi
\ifx \showDOI      \undefined \def \showDOI       #1{{\tt DOI:}\penalty0{#1}\ }
  \fi
\ifx \showISBNx    \undefined \def \showISBNx     #1{\unskip}     \fi
\ifx \showISBNxiii \undefined \def \showISBNxiii  #1{\unskip}     \fi
\ifx \showISSN     \undefined \def \showISSN      #1{\unskip}     \fi
\ifx \showLCCN     \undefined \def \showLCCN      #1{\unskip}     \fi
\ifx \shownote     \undefined \def \shownote      #1{#1}          \fi
\ifx \showarticletitle \undefined \def \showarticletitle #1{#1}   \fi
\ifx \showURL      \undefined \def \showURL       #1{#1}          \fi
\providecommand\bibfield[2]{#2}
\providecommand\bibinfo[2]{#2}
\providecommand\natexlab[1]{#1}
\providecommand\showeprint[2][]{arXiv:#2}

\bibitem[\protect\citeauthoryear{Agarwal, Balle, Gustavson, Joshi, and
  Palkar}{Agarwal et~al\mbox{.}}{1995}]%
        {Agarwal95athree-dimensional}
\bibfield{author}{\bibinfo{person}{Ramesh Agarwal}, \bibinfo{person}{Susanne
  Balle}, \bibinfo{person}{Fred~G. Gustavson}, \bibinfo{person}{Mahesh Joshi},
  {and} \bibinfo{person}{Prasad~V. Palkar}.} \bibinfo{year}{1995}\natexlab{}.
\newblock \showarticletitle{{A Three-Dimensional Approach to Parallel Matrix
  Multiplication}}.
\newblock \bibinfo{journal}{{\em IBM Journal of Research and Development\/}}
  \bibinfo{volume}{39} (\bibinfo{year}{1995}), \bibinfo{pages}{575--582}.
\newblock


\bibitem[\protect\citeauthoryear{Azad, Ballard, Bulu\c{c}, Demmel, Grigori,
  Schwartz, Toledo, and Williams}{Azad et~al\mbox{.}}{2016}]%
        {3d_algos}
\bibfield{author}{\bibinfo{person}{Ariful Azad}, \bibinfo{person}{Grey
  Ballard}, \bibinfo{person}{Aydin Bulu\c{c}}, \bibinfo{person}{James Demmel},
  \bibinfo{person}{Laura Grigori}, \bibinfo{person}{Oded Schwartz},
  \bibinfo{person}{Sivan Toledo}, {and} \bibinfo{person}{Samuel Williams}.}
  \bibinfo{year}{2016}\natexlab{}.
\newblock \showarticletitle{{Exploiting Multiple Levels of Parallelism in
  Sparse Matrix-Matrix Multiplication}}.
\newblock \bibinfo{journal}{{\em SIAM Journal on Scientific Computing\/}}
  \bibinfo{volume}{38}, \bibinfo{number}{6} (\bibinfo{year}{2016}),
  \bibinfo{pages}{C624--C651}.
\newblock


\bibitem[\protect\citeauthoryear{Ballard, Buluc, Demmel, Grigori, Lipshitz,
  Schwartz, and Toledo}{Ballard et~al\mbox{.}}{2013}]%
        {Ballard:2013:COP:2486159.2486196}
\bibfield{author}{\bibinfo{person}{Grey Ballard}, \bibinfo{person}{Aydin
  Buluc}, \bibinfo{person}{James Demmel}, \bibinfo{person}{Laura Grigori},
  \bibinfo{person}{Benjamin Lipshitz}, \bibinfo{person}{Oded Schwartz}, {and}
  \bibinfo{person}{Sivan Toledo}.} \bibinfo{year}{2013}\natexlab{}.
\newblock \showarticletitle{{Communication Optimal Parallel Multiplication of
  Sparse Random Matrices}}. In \bibinfo{booktitle}{{\em Proceedings of the
  Twenty-fifth Annual ACM Symposium on Parallelism in Algorithms and
  Architectures}} {\em (\bibinfo{series}{SPAA '13})}. \bibinfo{publisher}{ACM},
  \bibinfo{address}{New York, NY, USA}, \bibinfo{pages}{222--231}.
\newblock
\showISBNx{978-1-4503-1572-2}


\bibitem[\protect\citeauthoryear{Ballard, Druinsky, Knight, and
  Schwartz}{Ballard et~al\mbox{.}}{2016}]%
        {Ballard:2016:HPS:3012407.3015144}
\bibfield{author}{\bibinfo{person}{Grey Ballard}, \bibinfo{person}{Alex
  Druinsky}, \bibinfo{person}{Nicholas Knight}, {and} \bibinfo{person}{Oded
  Schwartz}.} \bibinfo{year}{2016}\natexlab{}.
\newblock \showarticletitle{{Hypergraph Partitioning for Sparse Matrix-Matrix
  Multiplication}}.
\newblock \bibinfo{journal}{{\em ACM Trans. Parallel Comput.\/}}
  \bibinfo{volume}{3}, \bibinfo{number}{3}, Article \bibinfo{articleno}{18}
  (\bibinfo{date}{Dec.} \bibinfo{year}{2016}), \bibinfo{numpages}{34}~pages.
\newblock
\showISSN{2329-4949}


\bibitem[\protect\citeauthoryear{Bethune}{Bethune}{2012}]%
        {dcse3}
\bibfield{author}{\bibinfo{person}{Iain Bethune}.}
  \bibinfo{year}{2012}\natexlab{}.
\newblock \bibinfo{booktitle}{{\em {CP2K} - Sparse Linear Algebra on 1000s of
  Cores}}.
\newblock \bibinfo{type}{{T}echnical {R}eport}.
  \bibinfo{address}{\url{http://www.hector.ac.uk/cse/
  distributedcse/reports/cp2k03/cp2k03.pdf}}.
\newblock


\bibitem[\protect\citeauthoryear{Borstnik, VandeVondele, Weber, and
  Hutter}{Borstnik et~al\mbox{.}}{2014}]%
        {dbcsr}
\bibfield{author}{\bibinfo{person}{Urban Borstnik}, \bibinfo{person}{Joost
  VandeVondele}, \bibinfo{person}{Valery Weber}, {and} \bibinfo{person}{Juerg
  Hutter}.} \bibinfo{year}{2014}\natexlab{}.
\newblock \showarticletitle{{Sparse Matrix Multiplication: The Distributed
  Block-Compressed Sparse Row Library}}.
\newblock \bibinfo{journal}{{\it Parallel Comput.}} \bibinfo{volume}{40},
  \bibinfo{number}{5-6} (\bibinfo{year}{2014}), \bibinfo{pages}{47--58}.
\newblock
\showISSN{0167-8191}


\bibitem[\protect\citeauthoryear{Bowler and Miyazaki}{Bowler and
  Miyazaki}{2012}]%
        {LS}
\bibfield{author}{\bibinfo{person}{David Bowler} {and}
  \bibinfo{person}{Tsuyoshi Miyazaki}.} \bibinfo{year}{2012}\natexlab{}.
\newblock \showarticletitle{{O(N) methods in electronic structure
  calculations}}.
\newblock \bibinfo{journal}{{\em Rep. Prog. Phys.\/}} \bibinfo{volume}{75},
  \bibinfo{number}{036503} (\bibinfo{year}{2012}).
\newblock


\bibitem[\protect\citeauthoryear{Buluc and Gilbert}{Buluc and Gilbert}{2008}]%
        {10.1109/ICPP.2008.45}
\bibfield{author}{\bibinfo{person}{Aydin Buluc} {and} \bibinfo{person}{John~R.
  Gilbert}.} \bibinfo{year}{2008}\natexlab{}.
\newblock \showarticletitle{{Challenges and Advances in Parallel Sparse
  Matrix-Matrix Multiplication}}.
\newblock \bibinfo{journal}{{\em 2008 37th International Conference on Parallel
  Processing (ICPP)\/}} (\bibinfo{year}{2008}), \bibinfo{pages}{503--510}.
\newblock
\showISSN{0190-3918}


\bibitem[\protect\citeauthoryear{Cannon}{Cannon}{1969}]%
        {cannon}
\bibfield{author}{\bibinfo{person}{Lynn~Elliot Cannon}.}
  \bibinfo{year}{1969}\natexlab{}.
\newblock {\em \bibinfo{title}{{A cellular computer to implement the Kalman
  Filter Algorithm}}}.
\newblock \bibinfo{thesistype}{Ph.D. Dissertation}. \bibinfo{school}{Montana
  State University}.
\newblock


\bibitem[\protect\citeauthoryear{Dekel, Nassimi, and Sahni}{Dekel
  et~al\mbox{.}}{1981}]%
        {3d_dense}
\bibfield{author}{\bibinfo{person}{Eliezer Dekel}, \bibinfo{person}{David
  Nassimi}, {and} \bibinfo{person}{Sartaj Sahni}.}
  \bibinfo{year}{1981}\natexlab{}.
\newblock \showarticletitle{{Parallel matrix and graph algorithms}}.
\newblock \bibinfo{journal}{{\em SIAM Journal on Scientific Computing\/}}
  \bibinfo{volume}{10}, \bibinfo{number}{4} (\bibinfo{year}{1981}),
  \bibinfo{pages}{657--675}.
\newblock


\bibitem[\protect\citeauthoryear{Gropp, Hoefler, Thakur, and Lusk}{Gropp
  et~al\mbox{.}}{2014}]%
        {UsingAdvancedMPI}
\bibfield{author}{\bibinfo{person}{William Gropp}, \bibinfo{person}{Torsten
  Hoefler}, \bibinfo{person}{Rajeev Thakur}, {and} \bibinfo{person}{Ewing
  Lusk}.} \bibinfo{year}{2014}\natexlab{}.
\newblock \bibinfo{booktitle}{{\em {Using Advanced MPI: Modern Features of the
  Message-Passing Interface}}}.
\newblock \bibinfo{publisher}{MIT Press}.
\newblock
\showISBNx{978-0262527637}


\bibitem[\protect\citeauthoryear{Gustavson}{Gustavson}{1978}]%
        {Gustavson:1978:TFA:355791.355796}
\bibfield{author}{\bibinfo{person}{Fred~G. Gustavson}.}
  \bibinfo{year}{1978}\natexlab{}.
\newblock \showarticletitle{{Two Fast Algorithms for Sparse Matrices:
  Multiplication and Permuted Transposition}}.
\newblock \bibinfo{journal}{{\em ACM Trans. Math. Softw.\/}}
  \bibinfo{volume}{4}, \bibinfo{number}{3} (\bibinfo{date}{Sept.}
  \bibinfo{year}{1978}), \bibinfo{pages}{250--269}.
\newblock
\showISSN{0098-3500}


\bibitem[\protect\citeauthoryear{Heinecke, Henry, Hutchinson, and
  Pabst}{Heinecke et~al\mbox{.}}{2016}]%
        {libxsmm}
\bibfield{author}{\bibinfo{person}{Alexander Heinecke}, \bibinfo{person}{Greg
  Henry}, \bibinfo{person}{Maxwell Hutchinson}, {and} \bibinfo{person}{Hans
  Pabst}.} \bibinfo{year}{2016}\natexlab{}.
\newblock \showarticletitle{{LIBXSMM: Accelerating Small Matrix Multiplications
  by Runtime Code Generation}}. In \bibinfo{booktitle}{{\em Proceedings of the
  International Conference for High Performance Computing, Networking, Storage
  and Analysis}} {\em (\bibinfo{series}{SC '16})}. \bibinfo{publisher}{IEEE
  Press}, \bibinfo{address}{Piscataway, NJ, USA}, Article
  \bibinfo{articleno}{84}, \bibinfo{numpages}{11}~pages.
\newblock
\showISBNx{978-1-4673-8815-3}


\bibitem[\protect\citeauthoryear{Higham}{Higham}{2008}]%
        {Higham:2008:FM}
\bibfield{author}{\bibinfo{person}{Nicholas~J. Higham}.}
  \bibinfo{year}{2008}\natexlab{}.
\newblock \bibinfo{booktitle}{{\em Functions of Matrices: {Theory} and
  Computation}}.
\newblock \bibinfo{publisher}{Society for Industrial and Applied Mathematics},
  \bibinfo{address}{Philadelphia, PA, USA}, \bibinfo{pages}{107--172}.
\newblock
\showISBNx{978-0-898716-46-7}


\bibitem[\protect\citeauthoryear{Hutter, Iannuzzi, Schiffmann, and
  VandeVondele}{Hutter et~al\mbox{.}}{2014}]%
        {cp2k}
\bibfield{author}{\bibinfo{person}{Juerg Hutter}, \bibinfo{person}{Marcella
  Iannuzzi}, \bibinfo{person}{Florian Schiffmann}, {and} \bibinfo{person}{Joost
  VandeVondele}.} \bibinfo{year}{2014}\natexlab{}.
\newblock \showarticletitle{{CP2K: Atomistic Simulations of Condensed Matter
  Systems}}.
\newblock \bibinfo{journal}{{\em Wiley Interdisciplinary Reviews: Computational
  Molecular Science\/}} \bibinfo{volume}{4}, \bibinfo{number}{1}
  (\bibinfo{year}{2014}), \bibinfo{pages}{15--25}.
\newblock
\showISSN{1759-0884}


\bibitem[\protect\citeauthoryear{Irony, Toledo, and Tiskin}{Irony
  et~al\mbox{.}}{2004}]%
        {IRONY20041017}
\bibfield{author}{\bibinfo{person}{Dror Irony}, \bibinfo{person}{Sivan Toledo},
  {and} \bibinfo{person}{Alexander Tiskin}.} \bibinfo{year}{2004}\natexlab{}.
\newblock \showarticletitle{{Communication lower bounds for distributed-memory
  matrix multiplication}}.
\newblock \bibinfo{journal}{{\em {Journal of Parallel and Distributed
  Computing}\/}} \bibinfo{volume}{64}, \bibinfo{number}{9}
  (\bibinfo{year}{2004}), \bibinfo{pages}{1017 -- 1026}.
\newblock
\showISSN{0743-7315}


\bibitem[\protect\citeauthoryear{McColl and Tiskin}{McColl and Tiskin}{1999}]%
        {McColl1999}
\bibfield{author}{\bibinfo{person}{William~F. McColl} {and}
  \bibinfo{person}{Alexander Tiskin}.} \bibinfo{year}{1999}\natexlab{}.
\newblock \showarticletitle{{Memory-Efficient Matrix Multiplication in the BSP
  Model}}.
\newblock \bibinfo{journal}{{\em {Algorithmica}\/}} \bibinfo{volume}{24},
  \bibinfo{number}{3} (\bibinfo{year}{1999}), \bibinfo{pages}{287--297}.
\newblock
\showISSN{1432-0541}


\bibitem[\protect\citeauthoryear{Rubensson and Rudberg}{Rubensson and
  Rudberg}{2016}]%
        {rubensson_rudberg}
\bibfield{author}{\bibinfo{person}{Emanuel Rubensson} {and}
  \bibinfo{person}{Elias Rudberg}.} \bibinfo{year}{2016}\natexlab{}.
\newblock \showarticletitle{{Locality-aware parallel block-sparse matrix-matrix
  multiplication using the Chunks and Tasks programming model}}.
\newblock \bibinfo{journal}{{\it Parallel Comput.}}  \bibinfo{volume}{57}
  (\bibinfo{year}{2016}), \bibinfo{pages}{87--106}.
\newblock


\bibitem[\protect\citeauthoryear{Schatz, Poulson, and van~de Geijn}{Schatz
  et~al\mbox{.}}{2012}]%
        {Schatz2012ParallelMM}
\bibfield{author}{\bibinfo{person}{Martin Schatz}, \bibinfo{person}{Jack
  Poulson}, {and} \bibinfo{person}{Robert van~de Geijn}.}
  \bibinfo{year}{2012}\natexlab{}.
\newblock \showarticletitle{{Parallel Matrix Multiplication: 2d and 3d Flame
  Working Note \#62}}.
\newblock


\bibitem[\protect\citeauthoryear{Sch{\"u}tt, Messmer, Hutter, and
  VandeVondele}{Sch{\"u}tt et~al\mbox{.}}{2015}]%
        {ole}
\bibfield{author}{\bibinfo{person}{Ole Sch{\"u}tt}, \bibinfo{person}{Peter
  Messmer}, \bibinfo{person}{Juerg Hutter}, {and} \bibinfo{person}{Joost
  VandeVondele}.} \bibinfo{year}{2015}\natexlab{}.
\newblock \showarticletitle{{GPU Accelerated Sparse Matrix Matrix
  Multiplication for Linear Scaling Density Functional Theory}}. In
  \bibinfo{booktitle}{{\em Electronic Structure Calculations on Graphics
  Processing Units}}. \bibinfo{publisher}{John Wiley and Sons}.
\newblock
\showISBNx{9781118661789}


\bibitem[\protect\citeauthoryear{Solomonik and Demmel}{Solomonik and
  Demmel}{2011}]%
        {25d}
\bibfield{author}{\bibinfo{person}{Edgar Solomonik} {and}
  \bibinfo{person}{James Demmel}.} \bibinfo{year}{2011}\natexlab{}.
\newblock \showarticletitle{{Communication-optimal parallel 2.5D matrix
  multiplication and LU factorization algorithms}}. In \bibinfo{booktitle}{{\em
  European Conference on Parallel Processing}}. Springer,
  \bibinfo{pages}{90--109}.
\newblock


\bibitem[\protect\citeauthoryear{{The CP2K Developers Group}}{{The CP2K
  Developers Group}}{2017}]%
        {cp2k_api}
\bibfield{author}{\bibinfo{person}{{The CP2K Developers Group}}.}
  \bibinfo{year}{2017}\natexlab{}.
\newblock \bibinfo{title}{{Overview of CP2K API}}.
\newblock
  \bibinfo{howpublished}{\url{https://apidoc.cp2k.org/index.html?dbcsr_api.html}}.
    (\bibinfo{year}{2017}).
\newblock


\bibitem[\protect\citeauthoryear{VandeVondele, Borstnik, and
  Hutter}{VandeVondele et~al\mbox{.}}{2012}]%
        {joost1M}
\bibfield{author}{\bibinfo{person}{Joost VandeVondele}, \bibinfo{person}{Urban
  Borstnik}, {and} \bibinfo{person}{Juerg Hutter}.}
  \bibinfo{year}{2012}\natexlab{}.
\newblock \showarticletitle{{Linear scaling self-consistent field calculations
  for millions of atoms in the condensed phase}}.
\newblock \bibinfo{journal}{{\em The Journal of Chemical Theory and
  Computation\/}} \bibinfo{volume}{8}, \bibinfo{number}{10}
  (\bibinfo{year}{2012}), \bibinfo{pages}{3565--3573}.
\newblock


\bibitem[\protect\citeauthoryear{Weber, Laino, Pozdneev, Fedulova, and
  Curioni}{Weber et~al\mbox{.}}{2015}]%
        {ibm}
\bibfield{author}{\bibinfo{person}{Valery Weber}, \bibinfo{person}{Teodoro
  Laino}, \bibinfo{person}{Alexander Pozdneev}, \bibinfo{person}{Irina
  Fedulova}, {and} \bibinfo{person}{Alessandro Curioni}.}
  \bibinfo{year}{2015}\natexlab{}.
\newblock \showarticletitle{{Semiempirical Molecular Dynamics (SEMD) I:
  Midpoint-Based Parallel Sparse Matrix-Matrix Multiplication Algorithm for
  Matrices with Decay}}.
\newblock \bibinfo{journal}{{\em The Journal of Chemical Theory and
  Computation\/}} \bibinfo{volume}{11}, \bibinfo{number}{7}
  (\bibinfo{year}{2015}), \bibinfo{pages}{3145--3152}.
\newblock


\end{thebibliography}

\end{document}